# Non-invasive single-shot recovery of point-spread function of a memory effect based scattering imaging system

Tengfei Wu,[1,2,3,*] Jonathan Dong,[3,4] Sylvain Gigan[3]

[1]*Sorbonne Université, CNRS, INSERM, Institut de la Vision, 17 Rue Moreau, 75012 Paris, France*
[2]*Université de Paris, SPPIN—Saints-Pères Paris Institute for the Neurosciences, CNRS, 75006 Paris, France*
[3]*Laboratoire Kastler Brossel, Sorbonne Université, École Normale Supérieure–Paris Sciences et Lettres (PSL) Research University, CNRS, Collège de France, 24 rue Lhomond, 75005 Paris, France*
[4]*Laboratoire de Physique de l'École Normale Supérieure, Université PSL, CNRS, Sorbonne Université, Université Paris-Diderot, Sorbonne Paris Cité, Paris, France*
*\*Corresponding author: tengfei.wu@parisdescartes.fr*

**Accessing the point-spread function (PSF) of a complex optical system is important for a variety of imaging applications. However, placing an invasive point source is often impractical, and estimating it blindly with multiple frames is slow and requires a complex non-linear optimization. Here, we introduce a simple single-shot method to non-invasively recover the accurate PSF of an isoplanatic imaging system, in the context of multiple light scattering. Our approach is based on the reconstruction of any unknown sparse hidden object using the autocorrelation imaging technique, followed by a deconvolution with a blur kernel derived from the statistics of a speckle pattern. A deconvolution on the camera image then retrieves the accurate PSF of the system, enabling further imaging applications. We demonstrate numerically and experimentally the effectiveness of this approach compared to previous deconvolution techniques. © 2020 Optical Society of America**

In complex imaging settings, optical scattering often prohibits the formation of a clear image and instead, only a complex speckle is obtained. To extract object information from the seemingly intractable speckle, a variety of techniques have been developed recently, several of which are based on the "Memory effect (ME)". This interesting physical phenomenon states that inherent angular correlations of scattered light exist for thin scattering layers [1,2]. Specifically, within the ME range, when we rotate the incident beam by a small angle, the structure of the resulted speckle pattern will not change but only translate over a certain distance. By exploiting the concept of ME, Katz et al. reported a single-shot convolution model to image hidden objects through scattering samples [3]. In this model, the camera image is regarded as a convolution of the image of hidden objects within the ME range and a shift-invariant point-spread function (PSF) of the scattering imaging system. Until now, this single-shot ME based convolution model has been widely used in many applications, for instance wide-field fluorescent microscopy [4], single-shot multispectral imaging [5], lens-less microscopy [6] or super-resolution deconvolution microscopy [7]. To solve the convolution model and retrieve the image of hidden objects, two strategies are commonly used. One of them is the speckle-correlation technique, in which the autocorrelation of the speckle pattern is calculated [3,8] to extract the Fourier amplitude of hidden object. Combining with a phase-retrieval process [9] or a high-order correlation of speckle pattern [10], its Fourier phase is recovered.

An alternative to phase retrieval is deconvolution: the PSF of a ME-based scattering imaging system is complex, but indeed deterministic. As long as the PSF can be obtained, a simple deconvolution process can in principle be used to solve the linear convolution model without using speckle-correlation. To implement deconvolution, a prerequisite is the calibration of intensity PSF of the scattering imaging system. Nowadays, a commonly-used way to access the PSF is using a point source and recording the speckle pattern. However, a natural point light-source in the biological sample is rarely available, and planting an artificial probe is impractical and invasive. It is possible to estimate the PSF of an optical system without using a point source, by reconstructing the pupil function (which contains more information than the intensity PSF) [11]. But recovering this pupil function requires multiple frames, which would sacrifice the temporal resolution and is impractical for e.g. dynamical samples. By using speckle-correlation and deconvolution, Lu et. al. estimated a PSF from a single-shot speckle pattern [12], but even though it enables the application of tracking objects, this method is less accurate as demonstrated in this work. Using stochastic source fluctuations, a nearly perfect PSF can be measured [13], nevertheless, it only works with isolated point-like sources (i.e. with negligible size compared to the diffraction limit).

Here, we demonstrate numerically and experimentally a simple single-shot method to non-invasively recover an accurate PSF of a ME based scattering imaging system, even with extended objects. Our method starts with the conventional autocorrelation imaging technique: the autocorrelation of a single-shot camera image and a phase-retrieval process are combined to retrieve a diffraction-limited image of any unknown hidden objects. As a major improvement over [12], we estimate the corresponding blur kernel of the diffraction-limited image and then retrieve a corrected image, approximating the true value, of the hidden objects from its diffraction-limited version. An accurate PSF is sequentially recovered by deconvolving the single-shot speckle pattern with the corrected object image. In addition to being straightforward to implement and possessing high temporal resolution (i.e. single-shot), this approach works with any unknown sparse object, which has the potential to being helpful for a wide range of applications, for instance, in biological imaging of fluorescent objects [4,8].

The principle and numerical simulations to validate this approach are presented in Fig. 1. In Fig. 1(a), the incoherent light carrying the information of a hidden object $O(r)$ [isolated points in Fig. 1(b)] propagates through an unknown scattering imaging system and forms a complex speckle pattern $I(r)$ where $r$ denotes a two-dimensional coordinate in the image plane. Considering that the object is within the ME range, the speckle pattern is the convolution of the object image and a deterministic PSF $S(r)$ of the scattering imaging system:

$$I(r) = O(r) * S(r) \quad (1)$$

where " $*$ " is a convolution operator. We consider the magnification of the system $M=1$ for simplicity in Eq. (1). The autocorrelation of speckle pattern satisfies the following identity:

$$I(r) \star I(r) = [O(r) \star O(r)] * [S(r) \star S(r)] \quad (2)$$

where " $\star$ " is a correlation operator. $S(r) \star S(r)$ can be modeled by a Bessel-based function [14] for a fully-developed speckle:

$$[S(r) \star S(r)] \propto \left| 2 \frac{J_1(\pi D r / \lambda v)}{\pi D r / \lambda v} \right|^2 \quad (3)$$

where $J(\cdot)$ denotes the first kind Bessel function, order one. $D$, $\lambda$ and $v$ are the pupil size, the wavelength and the image distance. Eq. (3) determines $S(r) \star S(r)$ is a sharply peaked function with a certain width (i.e. averaged speckle grain size) and influences the resolution of object autocorrelation $O(r) \star O(r)$. As a consequence, using the relationship of Eq. (2) and a basic phase-retrieval algorithm (See more details below in experimental description), although the image of the hidden objects can be retrieved from a single-shot speckle pattern, it is only a diffraction-limited version, which is related to the speckle grain size [Fig. 1(c)].

To access an accurate PSF of a scattering system, the image $O(r)$ should be first recovered as faithfully as possible, according to Eq. (1). One option is removing $S(r) \star S(r)$ on the object autocorrelation in Eq. (2) before the phase-retrieval process. However, we observed that deconvolution lowers the signal-to-noise ratio (SNR) of the object autocorrelation, always preventing a correct reconstruction from the phase-retrieval process.

To improve the technique, we propose to correct the blurred reconstruction *after* the phase-retrieval process. The diffraction-limited image $G(r)$ can be regarded as a convolution of the true image $O(r)$ and a blur kernel $\sigma(r)$:

$$G(r) = O(r) * \sigma(r) \quad (4)$$

and the blur kernel $\sigma(r)$ can be obtained by:

$$\sigma(r) = \mathcal{F}^{-1}\{|\mathcal{F}\{S(r)\}|\} \quad (5)$$

where $|\mathcal{F}\{\cdot\}|$ denotes the modulus of Fourier transform and $\mathcal{F}^{-1}\{\cdot\}$ is the inverse Fourier transform. Eq. (5) holds since $\sigma(r)$ is purely real and symmetrical, which is satisfied for an incoherent imaging system like in our case [15]. We give an additional justification for Eq. (5) by starting with the calculation of the Fourier transform of Eq. (2):

$$|\mathcal{F}\{I(r)\}|^2 = |\mathcal{F}\{O(r)\}|^2 \cdot |\mathcal{F}\{S(r)\}|^2 \quad (6)$$

where " $\cdot$ " is the multiplication. In the phase-retrieval process, the constraint on $G(r)$ in Fourier domain is the square root of Eq. (6), i.e. $|\mathcal{F}\{O(r)\}| \cdot |\mathcal{F}\{S(r)\}|$, which means the recovered diffraction-limited image from the phase-retrieval is low-pass filtered with a cut-off frequency determined by the width of $|\mathcal{F}\{S(r)\}|$, which can be deduced from the Bessel-function based autocorrelation model in Eq. (3). Therefore, the corresponding blur kernel $\sigma(r)$ in spatial domain can be obtained from Eq. (5).

Once $\sigma(r)$ is estimated, a faithful image $O'(r)$ [Fig. 1(d)] can be retrieved by using a simple deconvolution algorithm on Eq. (4). For implementing deconvolution, many pioneer methods have been proposed either in spatial domain or in Fourier domain. Here we employ the commonly-used Richardson–Lucy (RL) algorithm and 10 iterations are used to recover $O'(r)$. Finally, we use $O'(r)$ to deconvolve the camera image with 50 iterations of RL algorithm and an accurate PSF with full information is recovered in Fig. 1(g), which is highly correlated with the true PSF in Fig. 1(e). As a comparison, we try to use the diffraction-limited image $G(r)$ [Fig. 1(c)] to recover the PSF with the same process (as in [12]). It is clear that the recovered PSF, although being similar to the true one in structure, presents important artefacts, mainly caused by the loss of the information of speckle grain size, as can be seen in the zoom insets of Fig. 1(e), 1(f) and 1(g).

In the numerical examples above, we set the object distance $u$, the image distance $v$, the diameter of pupil aperture $D$, the wavelength $\lambda$, and the pixel size of the camera as 60cm, 12cm, 5.2mm, 633nm and 6.5μm respectively, corresponding to a speckle grain size of around 2.3 pixels. The angular spectrum method is used to simulate the light propagation between each optical element. By adjusting the pupil size to 2.9mm, we increase the speckle grain size to around 4 pixels and implement another group of simulation with an extended object "letter *P*". Fig. 1(h) shows the recovered diffraction-limited image by using the same basic phase-retrieval algorithm, and its inset is the estimated blur kernel $\sigma(r)$ in this case. The corresponding corrected image [Fig. 1(i)] is sequentially retrieved with 10 iterations of RL algorithm. 50 iterations are followed to recover the PSF by deconvolving the speckle pattern with the corrected image. The recovered PSF [Fig. 1(k)] still has a high correlation coefficient with the true value [Fig. 1(j)], and a zoom part is shown in the inset to compare the detail.

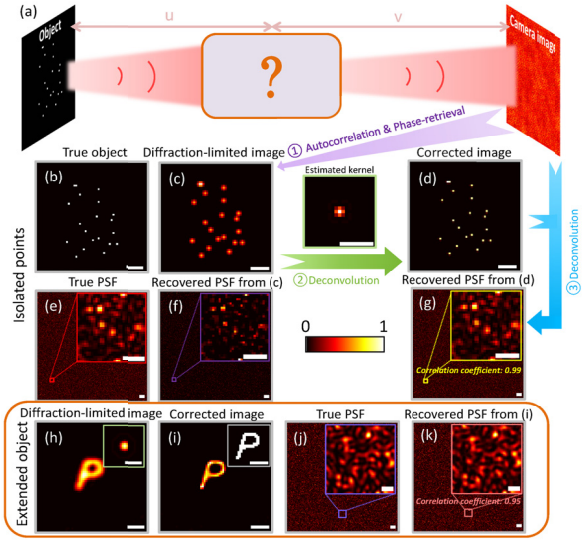

Fig. 1. Principle and numerical simulations of single-shot PSF recovery. (a) A scheme of a memory-effect based scattering image system. (b) True image of the isolated points. (c) Recovered diffraction-limited image. (d) Corrected image with an estimated kernel. (e) True PSF in the case of small speckle grain size. (f) Recovered PSF from diffraction-limited image. (g) Recovered PSF from corrected image (correlation coefficient: 0.99). (h) Recovered diffraction-limited image of an extended object (inset: estimated kernel). (i) Corrected image (inset: true image). (j) True PSF in the case of large speckle grain size. (k) Recovered PSF from corrected image (correlation coefficient: 0.95). Scale bar: 10 camera pixels in (b), (c), (d), (h), (i) and all the insets; 50 camera pixels in (e), (f), (g), (j) and (k).

Once an accurate PSF is recovered with our approach, it can be used to recover faithfully the image of other hidden objects. To demonstrate this numerically, we use another pattern of a relative complex structure "*OPTICS*" as the hidden object. The object is imaged on the camera in sequence through two scattering imaging systems with different pupil sizes and their PSFs in ideal condition (without noise) have been respectively estimated above in Fig. 1(g) and Fig. 1(k). For the case of large speckle grain size, we use the true PSF, the recovered PSF from corrected image and that from diffraction-limited image to respectively deconvolve the camera image in Fig. 2(a) with 200 iterations of RL algorithm. We notice that the deconvolution result with the PSF recovered by the proposed method in Fig. 2(c) is highly correlated with the reconstruction in Fig. 2(b) from the true PSF. Both of them can nearly recover perfectly the image of the hidden object. In comparison, the result in Fig. 2(d) is less resolved and moreover, due to the loss of the information of real speckle grain size, serious artefacts are present after the iterations.

Regarding the case of small speckle grain size, we add some Gaussian white noise, leading a SNR in the speckle to 25dB [Fig. 2(e)]. With the same procedure, two PSFs from corrected image and diffraction-limited image are measured. 300 iterations of RL algorithm are used to respectively deconvolve the camera image with the corresponding PSFs, and the results are shown in Fig. 2(f), 2(g) and 2(h). To more intuitively compare the reconstructions from different PSFs, we plot a row of the deconvolution results of three cases in Fig. 2(i). It is clear that in the same noisy condition, even though the reconstruction with PSF from the corrected image (red) has some deviations, comparing with the true value (green), it is still more faithful and better estimated than the one (blue) with the PSF recovered from diffraction-limited image.

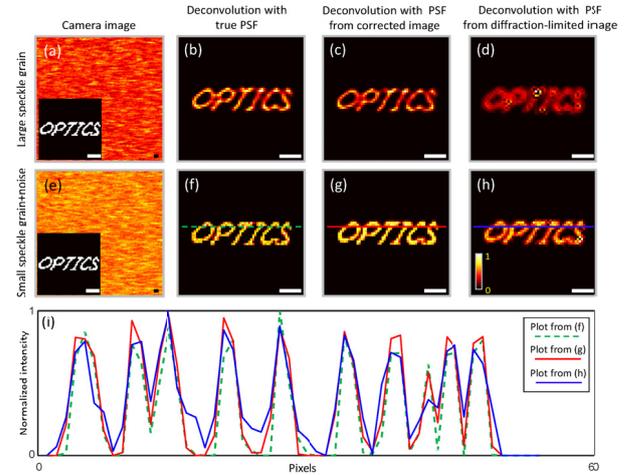

Fig. 2. Numerical results of deconvolution for other hidden objects with different PSFs. (a) Camera image with large speckle grain (inset: true image). Deconvolution result with (b) the true PSF, (c) the PSF recovered from corrected image and (d) recovered from diffraction-limited image. (e) to (h) are for the case of a small speckle grain size and in a noisy condition. The inset of (g) is in the ideal condition for comparison. (i) Comparison of a plotted line respectively from (f) (dashed green), (g) (red) and (h) (blue). Scale bar: 10 camera pixels in (b), (c), (d), (f), (g), (h) and all the insets; 50 camera pixels in (a) and (e).

We also present experimental results to demonstrate the interest of our approach in practice. Fig. 3(a) shows the optical setup. A ~350μm unknown object "digit 2" (Edmund Optics, 1951 USAF Target) is hidden ~60cm behind a thin scattering diffuser (Edmund Optics, Ground glass diffuser). A spatially incoherent LED (Thorlabs, M625L3) with a nominal wavelength 625nm and a bandwidth 18nm illuminates the hidden object and propagates through the diffuser. A narrow band-pass filter (Andover, 633FS02-50) is followed to ensure the speckle contrast. The illumination area on the diffuser is adjusted to 5.21mm by a contiguous diaphragm. A high-resolution digital sCMOS camera (Andor, ZYLA-5.5) is placed ~12cm in front of the diffuser to receive the speckle pattern.

Fig. 3(b) shows the raw speckle pattern of "digit 2" (inset: true image) received by the camera. Before calculating the autocorrelation, we spatially remove the envelope on the camera image by dividing it by its low-pass-filtered version. The Fourier amplitude of object is extracted by calculating the autocorrelation of the modified speckle pattern. A basic hybrid phase-retrieval algorithm is used to retrieve the diffraction-limited image of the hidden object from its Fourier amplitude. Specifically, we first employ the Hybrid Input-Output (HIO) algorithm with a decreasing $\beta$ factor from 2 to 0 with a step of 0.04 and run 50 iterations in each step. The result of HIO is then used as an initial guess of Error-Reduction algorithm for another 50 iterations to obtain the reconstruction. 200 random initial guesses are repeated in the phase-retrieval process and the mean square error (MSE) between the Fourier amplitude of each reconstruction and that from the speckle autocorrelation is used as a metric to monitor the convergence. The trial with lowest MSE value is selected as the optimized reconstruction for the following PSF recovery.

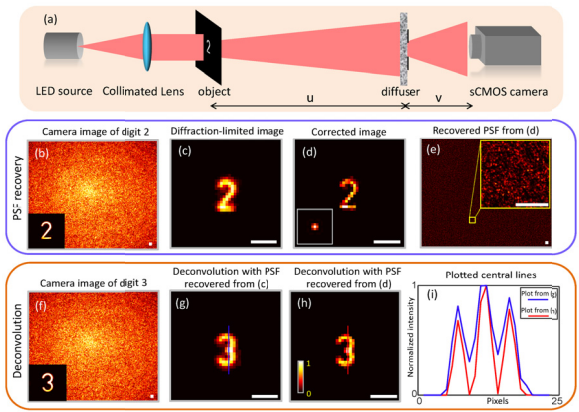

Fig. 3. Experimental results. (a) Experimental setup. (b) Camera image (inset for the hidden object). (c) Recovered diffraction-limited image. (d) Corrected image (inset: estimated kernel). (e) Recovered PSF from corrected image. (f) Camera image (inset: for another hidden object). Deconvolution result with recovered PSF from (g) diffraction-limited image and (h) corrected image. (i) Comparison of a plotted line from (g) in blue and (f) in red. Scale bar: 10 camera pixels in (c), (d), (g) and (h); 50 camera images in (b), (e), (e) inset and (f).

By using the estimated blur kernel [inset of Fig. 3(d)], a corrected image of "digit 2" [Fig. 3(d)] is retrieved from the diffraction-limited image [Fig. 3(c)] with 10 iterations of RL algorithm. Once the corrected image is retrieved, another 20 iterations of RL algorithm is used to deconvolve the camera image to recover the PSF [Fig. 3(e)] (Inset to see the detail). To experimentally validate our approach, we reconstruct another unknown object from the speckle pattern with the recovered PSF. We replace the original object "digit 2" with "digit 3" [inset of Fig. 3(f)] at around the same position, to keep the PSF unchanged. A raw speckle pattern [Fig. 3(f)] is then recorded, with which the image of "digit 3" is retrieved in Fig. 3(h) by using the previously-recovered PSF and 150 iterations of RL deconvolution algorithm. As a comparison, we use the diffraction-limited image [Fig. 3(c)] to recover the corresponding PSF, and the deconvolution result of image is obtained in Fig. 3(g). Either comparing directly the images or their plotting lines [Fig. 3(i)], it is clear that the reconstructed image in Fig. 3(h) is visually less affected by diffraction, since the recovered PSF from the corrected image is better estimated.

It is important to note that there is no claim of super-resolution using this technique. Even though we deconvolve by a blur kernel, there are many sources of noise (the experimental noise and the statistical noise with the autocorrelation imaging technique), on top of a non-linear optimization algorithm that is phase-retrieval. Nevertheless, our technique improves the fidelity of the PSF reconstruction and subsequent deconvolution techniques.

To summarize, we proposed a simple single-shot method to non-invasively recover a very accurate PSF of a ME based scattering imaging system. This approach relies on the use of autocorrelation, phase-retrieval algorithm and Bessel-function based model. It avoids implementing any point light-source behind or inside the scattering sample, instead, any incoherent sparse object should work for recovering the accurate PSF, as in fluorescent imaging scenarios in biomedical imaging [4,8]. Compared to the use of a point light-source to measure the PSF, using an extended object is less invasive and more straightforward, but the speckle contrast is also influenced by the sparsity $N$ of the object and scales as $1/\sqrt{N}$, which indicates that the lower the object sparsity, the lower the speckle contrast of the measured camera image [3,16]. Therefore, a trade-off between the object sparsity, total intensity and measurement noise should be considered. A high-resolution and sensitive camera with high dynamic range is therefore to be preferred (such as an sCMOS). An important extension of this work will be to extend it to finite ME scenario, i.e. to non-perfect isoplanetism. Another interesting extension may be the recovery of the complex pupil of a ME based imaging system with the accurate PSF and phase-retrieval process.

Code availability: we freely make available the source codes and experimental data to use by the scientific community, as shown in the Codes and Data file.

**Funding.** S.G. is funded by H2020 European Research Council (ERC, H2020, SMARTIES-724473).

**Acknowledgment**. S. G. acknowledges support from Institut Universitaire de France.

**Disclosures**. The authors declare no conflicts of interest.